\documentclass{ws-mpla}
\usepackage[super]{cite}
\usepackage{graphicx}
\usepackage{amsmath,amssymb}

\def\Journal#1#2#3#4{{#1} {\bf #2}, #3 (#4)}

\def\Sym{\em Symmetry}

\begin{document}


\catchline{}{}{}{}{}

\title{Introduction to the special issue\\Particle Dark Matter Candidates } 

\author{\footnotesize Maxim Yu. Khlopov${}^{a,b}$}
\address{${}^a$Laboratory of Astroparticle Physics and Cosmology 10, rue Alice Domon et Léonie Duquet, 75205 Paris Cedex 13, France\\
${}^b$Center for Cosmopartcile Physics “Cosmion” and National Research Nuclear University “MEPHI”
(Moscow State Engineering Physics Institute), Kashirskoe Shosse 31, Moscow 115409, Russia\\
khlopov@apc.univ-paris7.fr}

\maketitle


\begin{abstract}
A series of brief reviews collected in the present issue present various candidates for cosmological Dark Matter (DM) predicted by models of particle physics. The range from superlight axions to extended objects is covered. Though the possible list of candidates is far from complete it gives the flavor of the extensive field of Dark matter particle physics.
 
\end{abstract}

\ccode{PACS Nos.: include PACS Nos.}

The evidence for the existence of dark matter (DM) comes from gravitational lensing, measurements of galactic rotation curves, virial paradox in galactic clusters. These data indicate the existence of dominant nonluminous matter component, but doesn't prove its nonbaryonic nature. The latter comes from the measurements of large scale structures (LSS) and the anisotropy of cosmic microwave background (CMB), as well as from measurement of light element abundance in comparison with the predictions of the Standard Big Bang Nucleosynthesis (BBN). It means that dark matter particles co-exist with ordinary baryonic matter, dominate in the total matter density, and should be sufficiently stable to survive to the present time. These features move physics of dark matter beyond the Standard model (SM) of elementary particles.
\\

To form the Large Scale Structure of the Universe from initial density fluctuations as small as indicated by the measured CMB anisotropy, dark matter particles should decouple from plasma and radiation before the end of radiation dominated stage and provide growth of density fluctuations on matter dominated stage. Before recombination of hydrogen, baryonic matter, in which protons dominate, is in the state of electron-proton plasma, which experiences radiative pressure, converting at matter dominated stage growing modes of density fluctuations into the sound waves and thus preventing their effective growth. Such growth can start only after recombination, when the pressure of radiation doesn't affect neutral atomic gas. 
\\ 

In the absence of dark matter at the known law of growth of density fluctuations $\delta \rho/\rho$ at the matter dominated stage ($\delta \rho/\rho \propto a$, where $a$ is the scale factor) and known time of recombination, formation of LSS, corresponding to $\delta \rho/\rho \sim 1$ can take place only for the density fluctuations in the period of recombination by an order of magnitude larger, than measured by CMB anisotropy. In the presence of dark matter formation of LSS and measured anisotropy of CMB become compatible.
\\ 

Being one of the cornerstone of the modern cosmology, supported by cosmological and astrophysical observations, dark matter (DM) particles are still seem to be elusive from their direct experimental searches. In the context of General Relativity the very existence of these particles in our space time leads to their gravitational effect, while for the  description of the LSS formation at the observed level of CMB anisotropy sufficiently weak interaction of DM with plasma and radiation is needed. These features of DM candidates can be implemented in a rather wide class of particle models and of the corresponding cosmological scenarios. In particular, one should note that even ordinary strong (nuclear) interaction is sufficiently weak to provide decoupling of dark matter from plasma and radiation before recombination and thus explain formation of LSS and the observed level of CMB anisotropy,
\\

During last decades the mainstream of studies of dark matter candidates and their searches was concentrated around Weakly Interacting Massive Particles (WIMPs) predicted in supersymmetric models. The popularity of these candidates had several serious reasons. It naturally provided Cold Dark Matter scenario of LSS formation, favored in spite of some problems by observations. Weak interaction cross section of particles with the mass in the range of hundreds GeV naturally lead to their primordial abundance explaining the observed matter density of the Universe. Supersymmetric (SUSY) models naturally solving the problem of hierarchy for the electroweak symmetry scale and providing mechanism of electroweak symmetry breaking involved discrete or continuous symmetry supporting stability (or if broken, large lifetime) of the Lightest SuperSymmetric Particle (LSSP). Being neutral and having weak interaction with matter LSSP was a very attractive SUSY dark matter WIMP candidate. 
\\

However, the hope to discover supersymmetric particles with masses in the range of hundreds GeV at the LHC as well as to detect SUSY WIMPs in the direct underground searches for cosmic dark matter particles seems not to be proved. Though interpretation of positive results of dark matter searches in the DAMA/NaI and DAMA/LIBRA experiments is still not completely excluded, such interpretation is not supported by negative results of other experiments. It appeals to more extensive set of dark matter candidates and we discuss some representatives of this possible set in the present special issue.
\\

In fact, supersymmetry is only one attractive possible extension of the Standard model. Axion solution for the problem of CP violation in QCD is another important extension of the SM, offering solutions to its internal problems. Neutrinos are massless in the Standard model and the mechanism of neutrino mass generation is beyond it. One should also mention mirror world, restoring equivalence of left- and right- handed coordinate systems, as well as Grand Unification, unifying strong and electroweak interactions. 
\\

Not only extra gauge symmetries and additional sets of fundamental particles, but also the extra dimensions of the space-time can be involved in the extension of the Standard model. The extra dimensions can be compactified leaving only three infinite dimensions of our space.
They can be infinite, reducing our space and physics in it to a brane in a multi-dimensional space-time. 
\\

One can expect that the true measure for possible extensions of the Standard model can be found in a unique theoretical framework of some Theory of Everything and the string theory was considered as a possible basis for such a framework. However the model involving in the superhigh energy range not only high gauge symmetry, but also extra dimensions can hardly reduce to the Standard model in its low energy limit without many additional elements. 
\\

Even taken apart each of the popular extensions like supersymmetry or axion can predict many nontrivial cosmological consequences. Indeed, these examples demonstrate not only a variety of dark matter candidates, but also contain under some conditions such nontrivial cosmological predictions as antimatter domains in baryon asymmetrical Universe, or clouds of massive primordial black holes (see Ref.\citen{symmetry2} for review and references).  
\\

By construction mirror partners are strictly symmetric to their ordinary twins, having the same masses and couplings to the corresponding mirror bosons, as their ordinary partners\cite{LeeYang}. It seem to make mirror matter cosmology fully defined, being strictly symmetric to the cosmology of baryonic matter. But such strictly symmetric cosmological scenario with equal densities of mirror and ordinary matter and radiation is excluded by the constraints on the primordial helium abundance or local dark matter density, as well as, being equal to the ordinary baryonic density, mirror baryonic density is not sufficient to explain the observed density of dark matter\cite{Blin2,Khlopov:1989fj}. To be realistic, cosmology of mirror matter should lose its symmetry, either in the initial conditions\cite{zurab}, or in the parameters of mirror particles and their interaction. In the latter case asymmetric mirror matter becomes shadow matter with enormous variety of possible dark matter candidates \cite{Khlopov}.  
\\

In the contrast with the experimentally well proven Standard model, the lack of definiteness in the choice of a standard model for dark matter particle leads to necessity in the systematic analysis of all the physically motivated extensions of the Standard model with the aim to specify the features that can lead to such standard.
\\

The present issue can be considered as a small step in this direction. We
start with discussion of physical motivation for dark matter candidates in the extension of particle symmetry\cite{Khlopov} and consider dark matter candidates from Starobinsky Supergravity\cite{Addazi}. We give a brief review of axionic dark matter\cite{Qiaoli} and discuss probes for ultralight axion and scalar dark matter \cite{Stadnik}. We consider possible relationship between baryogenesis and dark matter in U(1) Extensions of the SM\cite{Nath}, give a brief review of interacting dark matter and dark radiation\cite{Tang} and of heavy right-handed neutrino dark matter in left-right models\cite{Zhang}. We finally discuss extended micro objects and dark atoms as dark matter particles\cite{Rubin}. Being sufficiently representative this list of possible candidates is far from complete and we hope that discussion of the problem of particle dark matter candidates will continue. In particular, more extensive discussion of this problem will be presented in the coming books Refs. \citen{BookReview,BookDADW}. 
 
\section*{Acknowledgments}
I am grateful to all the authors of the present issue for their valuable contributions. The work was performed within the framework of the Center FRPP supported
by MEPhI Academic Excellence Project (contract 02.03.21.0005,
27.08.2013).


\end{document}